\documentclass[12pt, a4paper]{article}
\usepackage{geometry}
\usepackage[utf8]{inputenc}

\usepackage[T1]{fontenc} 
\usepackage{graphicx,epstopdf} 

\usepackage{bm}
\usepackage{secdot}
\usepackage{mathptmx}
\usepackage{float}
\usepackage{soul}
\usepackage{authblk}

\RequirePackage{graphicx}
\RequirePackage{amsmath}
\RequirePackage{amssymb}

\title{Dissipative Unified Dark Fluid Model}

\author{Esraa Elkhateeb} 
\affil{\small Physics Department, Faculty of Science, Ain Shams University., Abbassia 11566, Cairo, Egypt}

\begin{document}
\date{}

\maketitle
\begin{abstract} 
{We consider a unified barotropic dark fluid model with dissipation. Our fluid asymptotes between two power laws and so can interpolate between the dust and dark energy equations of state at early and late times. The dissipative part is a bulk viscous part with constant viscosity coefficient. The model is analyzed using the phase space methodology which helps to understand the dynamical behavior of the model in a robust manner without reference to the system solution. The parameters of the model are constrained through many observational constraints. The model is tested through many physical and observational tests. We first considered the model independent $Om(z)$ test. Results for $Om(z)$ are plotted against the BAO data for this quantity from different authors, which shows that the model is consistent with the data points for the full redshift range. The $\chi^2$ statistics results in the value of ${\chi^2}_{red}=0.30$ with a $p-$value of $0.95$. The Hubble parameter equation is solved numerically and results are plotted against the recent set of Hubble data. The ${\chi^2}$ test with the Hubble data resulted in the ${\chi^2}_{red}$ value of $0.672$ with a $p-$value of $0.94$. The distance modulus at different values of redshift is calculated numerically and results are compared to the newest set of SNe Ia data, the Pantheon Sample. We obtained a ${\chi^2}_{red}$ value of $1.04$ with a $p-$value of $0.17$. These results show that our model is efficiently consistent with observations. The model expectations for the evolution of the universe are also studied by testing the evolution of the deceleration parameter, the density of the universe, and the effective equation of state parameter of the model and of its underlying dark energy candidate. The value of the present day viscosity coefficient of the cosmic fluid, $\zeta_0$, is estimated. It is found to be $8 \times 10^6 Pa$ $s$. We argue that this model is able to explain the behavior of the universe evolution.}
\end{abstract}



\section{Introduction}
In the standard model of cosmology \cite{wie, peeb}, the dark energy (DE) component, which based on observations constitutes about $72\%$ of the universe \cite{jons}, is represented by the cosmological constant $\Lambda$. This component is responsible for the current accelerated expansion of the universe. The second leading component in this model is a non-relativistic matter component which interacts only gravitationally and is known as cold dark matter (CDM). Other components are baryons, photons, and neutrinos. Even though this model, the $\Lambda$$CDM$ model, agrees well with observations, it faces some debatable issues. For instance, the huge gap between the observed value of the vacuum energy density and that expected from the quantum field theory is $120$ orders of magnitude. Another problem is the coincidence of the order of magnitude between the measured vacuum energy density and the matter energy density for the present time, although the former is presumed to be constant in time while the latter decreases with it. This leads cosmologists to visualize other scenarios.

One of these scenarios is the dynamical DE scenario, in which the constant DE component is replaced by a dynamical one. The most natural approach is to consider a dynamical DE equation of state (EoS) evolving with cosmic expansion, where the EoS can be parametrized in different ways \cite{lin, che, sche, mam, mam1, sendra}.

However, there is still no direct detection for the DM and the DE proposed by the $\Lambda$$CDM$ model. In fact, It is not established if they are really independent entities. The possibility of the unified scene of DE and DM has attracted a lot of interest. The class of models that unifies these two dark sectors is often referred to as quartessence \cite{RRRR, RRRM}. Prototypes of these models are the Chaplygin gas (CG) model and its generalizations (GCG) \cite{kamen, baf, sar, din, gor}. Among other unification scenarios, there are also the tachyonic field \cite{gib, pad, sen} and the condensate cosmology \cite{bas}.

One of the special interesting class of models is the barotropic fluid models. Barotropic fluids are characterized by an EoS for which the pressure is an explicit function of the density, $p=f(\rho)$. Barotropic models are considered by many authors either as unified models, of which CG and GCG are again examples, or as models for DE alone. Linder and Scherrer \cite{linder} studied extensively the general properties of the barotropic DE models and how they can be distinct from quintessence. $\check{\texttt{S}}$tefan$\check{\texttt{c}}$i$\acute{\texttt{c}}$ \cite{stef, stef1} considered a barotropic DE equation of state which is some sort of expansion around the cosmological constant EoS. Different forms of barotropic DE equations of state are also studied by Nojiri and Odintsov \cite{nojiri}.

Unified models are promising. Their idea has sprung historically from the cosmological properties of Chaplygin gas, an exotic fluid that has the dynamical property to behave as dark matter in the early time and as dark energy in the late time. However, although CG models succeeded on the level of cosmic background dynamics\cite{fabrs, fabrs1}, they fail to produce the matter power spectrum. Sandvik et al.\cite{sandv} showed that on the perturbation level the matter power spectrum data strongly constrain the parameters of the generalized Chaplygin gas models leaving narrow room of allowed such models indistinguishable from the $\Lambda$CDM model. On the other hand, Reis et al.\cite{reisetal} showed that the ad hoc inclusion of entropy perturbations to the GCG models enlarges the parameter space, so that for a wide range of the parameter space, the instabilities and oscillations disappear and results are compatible with large scale structure (LSS) and cosmic microwave background (CMB) observations. Many authors, e.g., Hip$\grave{o}$lito-Ricaldi et al.\cite{hopo, hopoetal}, Fabris et al.\cite{fabrisetal}, and Borges et al.\cite{borges}, then considered viscosity as a natural candidate for intrinsic entropy perturbation. The authors showed that the perturbation dynamics for their models avoid short scale oscillations or instabilities. They also showed how the viscous dark fluid models are well competitive with $\Lambda$CDM model.

Dissipative cosmology was studied as early as 1967 when Zel'dovich\cite{zel}, on calculating the universe's entropy, showed that the present specific entropy of the universe can be calculated by considering the action of the dissipative processes in the early universe. The most simple model of the viscous universe was proposed by Padmanabhan and Chitre \cite{padm} in 1987. They considered a universe model dominated by dust with constant viscosity coefficient. They came to the conclusion that viscosity can be neglected at early times, while at late times it causes the universe to enter a late inflationary era with exponentially accelerated expansion. Later on, a lot of work was done considering viscous cosmology. Fabris et al. \cite{fabris} studied the possibility that the present accelerated expansion of the universe is driven by a viscous fluid. Their fluid was controlled by Eckart's formalism for bulk viscosity. They showed that although their model leads to the same results of GCG model for some choices of the parameters, it shows an absence of instabilities in the power spectrum for any choice of the parameter $\nu$ of the viscosity coefficient, $\left( \zeta(\rho)=\zeta_0 \rho^\nu \right)$. They also showed that their model has a more normal situation since the viscosity grows with density.  Many other authors, e.g., Avelino and Nucamendi \cite{art} and Li and Barrow \cite{lib}, considered the possibility that the present acceleration of the universe is driven by bulk viscous pressure.

Brevik and Gr{\o}n \cite{brev} studied the effect of different types of viscosity on the decay of anisotropy of the universe. The authors came to the conclusion that the existence of viscosity tends to smooth out anisotropies in the universe. Another comprehensive study for the viscous early and late universe is made by Brevik et al. \cite{brevetal}. The authors studied in detail the effect of viscosity on various inflationary observables. They could describe uniquely inflationary and current cosmic acceleration in the framework of viscous cosmology. One of the important results of their work is the conclusion that the magnitude of the bulk viscosity constrained by observations may be sufficient to drive the cosmic fluid from quintessence to phantom region.

A general viscous isotropic flat Friedmann universe was studied by Norman and Brevik \cite{norman}. An important result of their work is the calculation of the current viscosity coefficient constrained by Hubble parameter observations for the considered cosmological fluid to a non-zero value. Many authors, e.g., Velten and Schwarz \cite{velt}, Wang and Meng \cite{wang}, Brevik \cite{brvi}, and Sasidharan and Mathew \cite{sasi} also came to the same result.

The present work aims to study a unified viscous dark fluid model. We build on the idea of $\check{\texttt{S}}$tefan$\check{\texttt{c}}$i$\acute{\texttt{c}}$ \cite{stef, stef1} and Nojiri and Odintsov \cite{nojiri}. The EoS of our fluid is a generalization of the EoS proposed in their work. We studied the background dynamics of this model as a perfect fluid in a previous work \cite{esra} and showed that it agrees very well with the supernovae Ia data and other observations such as today's value of deceleration parameter $q_0$. In this work, we extend the previous work by adding dissipative effects as a natural candidate of a real fluid. These dissipative effects are added in the form of bulk viscosity. Shear viscosity is considered negligible, as observations indicated that the universe is isotropic.
 
The manuscript is organized as follows: Section \ref{dynamics} introduces the basic dynamics of the bulk viscous cosmology and applying it to the unified model studied in this work. In section \ref{phase} we analyze the model using the phase space methodology. Section \ref{cosmpar} constrains the parameters of the model using cosmological observations. In section \ref{test} we confront the model with many basic physical and observational tests and study the model expectations for the universe evolution. In section \ref{conc} we present our final conclusions.

\section{Dynamics of the Model} \label{dynamics}
In the standard FRW cosmology, the flat homogeneous and isotropic universe is described by the metric
\begin{equation}
ds^2=dt^2-a^2(t) \delta_{ij} dx^i dx^j
\label{metric}
\end{equation}
where we consider units with $c=1$. The energy momentum tensor for the fluid is
\begin{equation}
T_{\mu\nu}=\rho U_\mu U_\nu+\left(p-\theta \zeta \right)h_{\mu\nu}
\label{tensor}
\end{equation}
where $\rho$ is the energy density of the cosmic fluid, $\theta=3 H$ is the volume expansion rate of the fluid with $H=\frac{\dot{a}}{a}$ is the Hubble parameter, $\zeta=\zeta(\rho)$ is the coefficient of bulk viscosity that arises in the fluid and is restricted to be positive, and $h_{\mu\nu}=U_\mu U_\nu-g_{\mu\nu}$ is the projection tensor to the $3-$space orthogonal to the fluid element, where in comoving coordinates the four-velocity $U_\mu={\delta^0}_\mu$.

With the metric (\ref{metric}), Einstein's equation readily leads to Friedman equations:
\begin{align}
  \frac{\dot{a}^2}{a^2}=\frac{8 \pi G}{3}\rho \label{dot}\\
	\frac{\ddot{a}}{a}=-\frac{4 \pi G}{3}(\rho+3P) \label{ddot}
\end{align}
where the fluid pressure $P$ is considered as a barotropic pressure with dissipation
\begin{equation}
  P=p-\theta \zeta(\rho)
\label{pres}
\end{equation}

Considering units with $8 \pi G=1$, Relations (\ref{dot}) and (\ref{ddot}) reduce to
\begin{align}
H^2=\frac{1}{3} \rho	\label{hsq} \\
H^2+\dot{H}=-\frac{1}{6}(\rho+3P) \label{hddot}  
\end{align}

The conservation equation ${{T_\nu}^\mu};\mu=0$ gives
\begin{equation}
\dot{\rho}+3 \frac{\dot{a}}{a} (\rho + P)=0       
\label{cons}
\end{equation}
This relation can be written in terms of Hubble parameter as
\begin{equation}
\dot{H}+\frac{1}{2} (\rho + P) = 0       
 \label{rodot}
\end{equation}

To close the system of equations, the Eos of the cosmic fluid must be considered. Linder and Scherrer \cite{linder} studied the barotropic fluid models extensively. They showed that these models are attractive and promising. Nojiri and Odintsov \cite{nojiri} studied the dynamical DE models for which the DE is described by a barotropic fluid with the general EoS
\begin{equation}
  p=-\rho - f(\rho)
\label{stefeq}
\end{equation}
A later study by $\check{\texttt{S}}$tefan$\check{\texttt{c}}$i$\acute{\texttt{c}}$ \cite{stef, stef1} considered such EoS for DE with $f(\rho) =  A \rho^\alpha$, where $A$ and $\alpha$ are real parameters and $\alpha \neq 1$. Nojiri et al. \cite{nojiri1, nojiri2} also studied the future singularities associated with this model.

In the present article we build on the work of the previous authors, although we follow the unified scene. We consider a barotropic pressure of the form
\begin{equation}
  p=-\rho + \frac{\gamma \rho^n}{1+\delta \rho^m}
\label{van1}
\end{equation}
where $\gamma$, $\delta$, $n$, and $m$ are free parameters. The background dynamics of such a perfect fluid were studied in previous work \cite{esra}. It was shown that this form of barotropic pressure has the advantage of interpolation between different powers for the density, whence allows for smooth phase transitions during the universe evolution. It was also shown that this EoS has the advantage of being general for describing DE, so that it enables the cosmological constant as a special case. In this work we add dissipative effects as a natural candidate of the real fluid. These dissipative effects are described by the bulk viscosity. The coefficient of bulk viscosity is considered to take the form
\begin{equation}
\zeta(\rho) = \zeta_0 \rho^\nu
\label{vis}
\end{equation}
where $\zeta_0$ and $\nu$ are constants. However, we will focus on the simple ansatz of $\nu=0$ so that we have a constant bulk viscosity coefficient $\zeta(\rho) = \zeta_0$.

\section{Cosmological Model Evolution in View of the Theory of Dynamical System} \label{phase}
The theory of dynamical systems \cite{brin, lyn}, is originated to study the long term behavior of evolving systems. In this theory, the phase space is a multidimensional space in which each dimension represents one degree of freedom of the dynamical system. A dynamical system can, in general, be described by an autonomous system of differential equations $\bf{\dot{x}}=\bf{f}\left(\bf{x}\right)$, where in the n-dimensional system the vectors $\textbf{x}$ and $\textbf{f}\left(\textbf{x}\right)$ are given by $\textbf{x}=\left(x_1, x_2, .. x_n\right)$ and $\textbf{f}\left(\textbf{x}\right)=\left(f_1\left(\textbf{x}\right), f_2\left(\textbf{x}\right), .. f_n\left(\textbf{x}\right)\right)$. Many important features of the motion can be guessed from the phase portrait of the dynamical system without solving the equations of motion in detail.

The application of the theory of dynamical systems to cosmological scenarios is very powerful \cite{col, cop, rich, aawad, aawad1}. In most of the cosmological models, the cosmological equations, in spite of the difficulty of their analytical solution, may have many solution branches due to different initial conditions. The application of the phase space method allows the extraction of essential information about different solutions of the system, and about the evolution of the cosmological model and its asymptotic dynamics. Accordingly, expectations about the origin and fate of the universe can be extracted. In addition, it allows the discussion of the stability of the model solution (a comprehensive discussion about the application of the theory of dynamical systems to FLRW cosmology and consequences of the existence of fixed points on finite time singularities of different types is found in the work of Awad\cite{aawad}).

Upon applying the theory to our model, the combination of eqns. (\ref{van1}) and (\ref{rodot}) yields
\begin{equation}
\dot{H}=-\frac{1}{2}\left(\frac{\alpha H^r}{1+\beta H^s}- 3 \zeta_0 H\right)
\label{dhdt1}
\end{equation}
where $\alpha$, $\beta$, $r$, and $s$ are constants related to those of (\ref{van1}) through the relations
\begin{equation}
r=2 n \; ; \; \; \; s=2 m \; ; \; \; \; \alpha=3^n \gamma \; ; \; \; \; and \; \; \; \beta=3^m \delta
\label{const}
\end{equation} 

As mentioned in García-Salcedo \cite{rich}, one of the unwritten rules that one has to follow in choosing the variables of the phase space is that they should be dimensionless. Note that in (\ref{dhdt1}), $(\alpha H^r)$ has the dimensions of Hubble parameter squared, consequently $[\alpha]$=$[H^{-(r-2)}]$. On the other hand, $(\beta H^s)$ is dimensionless, which means that $[\beta]$=$[H^{-s}]$. Note also that since we consider units with $8 \pi G = c =1$, the dimensions of $\zeta_0$ is the same as Hubble parameter,  $[\zeta_0]=[H_0]$.  Accordingly, we define the dimensionless parameters
\begin{equation}
h=\beta^{1/s} H ;   \, \; \, \, \; \, \tau=\alpha \beta^{(1-r)/s} t
\label{dimles}
\end{equation}
Eq. (\ref{dhdt1}) then takes the form
\begin{equation}
\frac{d h}{d \tau}=-\frac{1}{2} \left(\frac{h^{r}}{1+h^s} - \widetilde{\zeta_0} h\right)
\label{htau}
\end{equation}
where 
\begin{equation}
\widetilde{\zeta_0}=\frac {3 \beta^{(r-1)/s}}{\alpha} \zeta_0
\label{tzet}
\end{equation}
Our model must mimic dust at large $h$ and dark energy at small $h$, so that $dh/d\tau$ has to be proportional to $h^2$ for large $h$, and tends to zero for small $h$. As a result, the parameter $r$ must be positive. Restricting $s$ also to be positive, (\ref{htau}) tends asymptotically to the following form for large $h$
\begin{equation}
\frac{d h}{d \tau} \sim -\frac{1}{2} \left( h^{r-s}-\widetilde{\zeta_0} h \right)
\label{asymp1}
\end{equation}
If $\widetilde{\zeta_0}$ is smaller or comparable to $h$, the above equation mimics dust for $(r-s) > 1$, otherwise the term linear in $h$ dominates. Under these conditions, eq (\ref{asymp1}) reduces to 
\begin{equation}
\frac{d h}{d \tau} \sim -\frac{1}{2} h^{r-s}
\label{asymp}
\end{equation}
 Comparing this with Friedmann eqn for dust, namely 
\begin{equation}
\frac{d h}{d \tau}=-\frac{3 \beta^{(r-2)/s}}{2 \alpha} h^2
\label{dust}
\end{equation}
We get
\begin{equation}
r-s=2 \; ; \; \; \; and \; \; \; \; \frac{\alpha}{\beta}=3
\label{par}
\end{equation}
As a result
\begin{equation}
\widetilde{\zeta_0}=\frac {\zeta_0}{\widetilde{\alpha}} 
\label{tzeta}
\end{equation}
where $\widetilde{\alpha}=\left(\frac{\alpha}{3}\right)^{(-1/s)}$ has the dimensions of Hubble parameter, $[\widetilde{\alpha}]=[H]$. Apply (\ref{par}) to (\ref{htau}) we get
\begin{equation}
\frac{d h}{d \tau}=-\frac{1}{2} \left(\frac{h^{s+2}}{1+h^s} - \widetilde{\zeta_0} h\right)
\label{htau1}
\end{equation}
The effective equation of state parameter, $\omega_{eff}= P / \rho$, is calculated using  (\ref{pres}) and (\ref{van1}). This gives
\begin{equation}
  \omega_{eff}=-1 + \frac{h^s}{1+h^s}- \frac{\widetilde{\zeta_0}}{h} 
\label{effwh}
\end{equation}
Accordingly, relation (\ref{htau}) can also be written as
\begin{equation}
  \frac{d h}{d \tau}=\frac{-1}{2} \left( 1 + \omega_{eff}\left(\tau\right) \right) h^2  
\label{phas}
\end{equation}
Relations (\ref{htau}) and (\ref{phas}) trace the trajectory of the system in its phase space.
	\begin{figure} [ht]
	\centering
		\vspace{-1cm}
		\includegraphics[width=11.5cm, height=9cm] {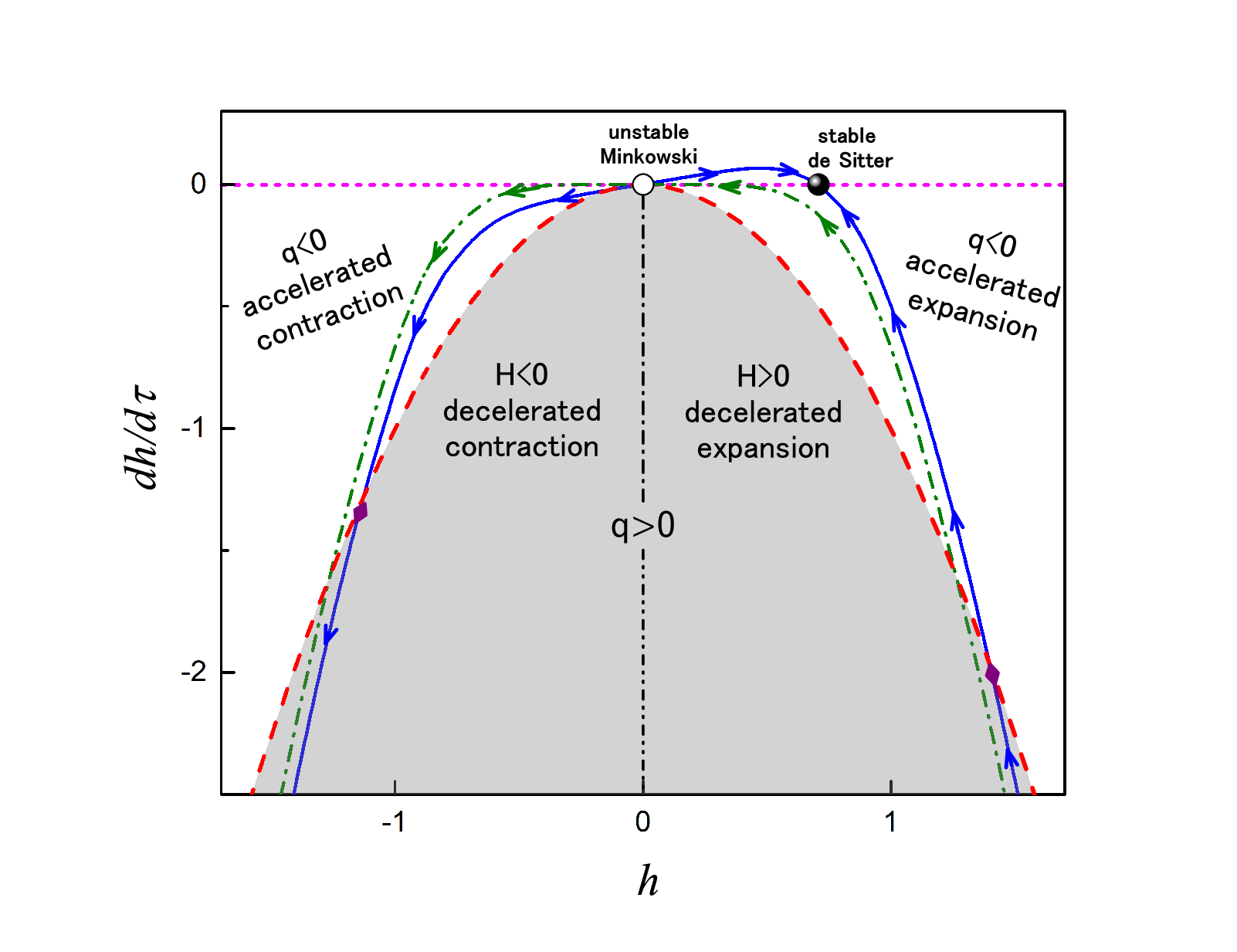}
  	\vspace*{-.5cm}
	\caption{\scriptsize{Phase diagram of our model, solid blue curve, for $\zeta(\rho)=\zeta_0$, compared to previous work with $\zeta(\rho)=0$ \cite{esra}, dash-dot green curve. The value of $\widetilde{\zeta_0}$ is taken as $0.1$. The zero acceleration limit, dash red curve, separates the deceleration region, shaded area, from acceleration regions. The two purple diamonds show the transition between decelerated and accelerated expansion regions.}}
	\label {fig:phase}
\end{figure}

 Fig.\ref{fig:phase} shows the phase portrait for our viscus fluid model, $\zeta(\rho)=\zeta_0$ against the corresponding perfect fluid model of the previous study \cite{esra}, $\zeta(\rho)=0$. The value of $\widetilde{\zeta_0}$ is chosen arbitrarily as $0.1$. In the following section, we will constrain all of the model parameters through observations.The red boundary represents the zero acceleration limit. Recall that the deceleration parameter is defined by
\begin{equation}
q(z) = -\frac{\ddot{a}}{a} \frac{1}{H^2}
\label{dec1}
\end{equation}
the shaded area then represents the deceleration phase of the evolution where $q > 0$. Inside this region, expansion takes place if $H > 0$, while contraction takes place if $H < 0$. Outside this region we find the acceleration phase, where $q < 0$, and where expansion accelerated if $H > 0$ while contraction accelerated if $H < 0$.  

We can see from the Fig. that the bulk viscous pressure trajectory characterized by two fixed points. One of these points is the null solution where $(h, \frac{d h}{d \tau})=(0,0)$. This is an unstable fixed point, (although it is a semi-stable point with respect to the perfect fluid model). It represents an unstable transient Minkowskian empty space since the matter density drops to zero asymptotically, $h \rightarrow 0$. To reach this point, the trajectory exhibits a positive slope, which, due to (\ref{phas}), means that $\omega_{eff} < -1$. The universe then crosses the phantom line in this region.

The other fixed point, $(h,0)$, is a stable future attractor. Referring to relation (\ref{phas}), since $h \neq 0$ at this point then $\frac{d h}{d \tau}=0$ when $\omega_{eff}=-1$ . Accordingly, this point represents a stable de Sitter universe dominated by dark energy. It is the cosmological constant asymptote of the model. 

As fixed points are stationary it follows that if the dynamical system starts at a fixed point it'll remain there forever, these are de Sitter cosmologies. It is thus clear that we have a multi-branch solution, so that different initial conditions lead to different solutions. One of these solutions exists in the negative $H$ patch. In fact, this does not match our universe. We know from observations that $H$ is positive, we live in an expanding universe. In the other solution, the system evolves between the two fixed points. If the system starts somewhere between two fixed points it'll evolve smoothly without any singularities. However, although this solution exists in the positive $H$ branch, such universe is again not ours. It has properties which contradict observations, such as the positivity of $\dot H$ and the nonexistence of the transition phases. A third solution evolves in the positive $H$ patch until it reaches the fixed de Sitter point. The physics of the evolution of this universe matches ours. It evolves from a Big Bang singularity to a de Sitter space. It evolves from decelerated to accelerated expansion.  And since $f(h)=\frac{d h}{d \tau}$ is continuous and differentiable, it is practically evolving to a de Sitter space free from future time singularities \cite{aawad}, which due to the classification of Nojiri et al. \cite{nojiri1} are
\begin{itemize}
	\item Type $I$ (Big Rip) singularity: where all of the scale factor, the energy density and the pressure of the universe diverge ($\left|P\right| \rightarrow \infty$, $a \rightarrow \infty$ and $\rho \rightarrow \infty$) in a finite time $t \rightarrow t_s$.
	\item Type $II$ (Sudden) singularity: where only the pressure of the universe diverges $\left|P\right| \rightarrow \infty$ in a finite time, as $t \rightarrow t_s$, while $a \rightarrow a_s$ and $\rho \rightarrow \rho_s$. 
	\item Type $III$ singularity: where both the pressure and the effective energy density diverge ($\left|P\right| \rightarrow \infty$ and $\rho \rightarrow \infty$) in a finite time $t \rightarrow t_s$ while $a \rightarrow a_s$. 
\end{itemize}
 
Accordingly, this third solution is adopted.

\section{Cosmological Parameters of the Model} \label{cosmpar}
Our model is a unified dark fluid model which asymptotes between two power laws that describe the two phases of dust and DE. This enables the cosmic fluid to interpolate smoothly between dust at the early time and dark energy at the late time. Specifically, at the late time we have a more general EoS for DE which enables cosmological constant as a special case. In this section we are going to constrain the model parameters. In Eq.(\ref{par}) two of the parameters of the EoS were constrained. The remaining parameters will be constrained using cosmological observations.
 
\subsection{Constraining the Bulk Viscosity Coefficient with $q_0$} \label{decepar}
The present day value of deceleration parameter (DP), $q_0$, is one of the most important cosmological parameters. In his (1970) paper, Alan Sandage \cite{sand} defined the observational cosmology as the search for two parameters: the Hubble parameter $H_0$ and the deceleration parameter $q_0$.

DP is defined as a dimensionless dynamical parameter given by (\ref{dec1}). Now using 
\begin{equation}
\frac{\ddot{a}}{a}=H^2+\dot{H}
\label{dec2}
\end{equation}
relation (\ref{dec1}) will take the form
\begin{equation}
q(z) = -1 - \frac{\dot{H}(z)}{H^2(z)}
\label{dec3}
\end{equation}
The relation of $\dot{H}$ for our model is found by using Eq. (\ref{par}) in (\ref{dhdt1}). This gives
\begin{equation}
\dot{H}=-\frac{1}{2}\left(\frac{\alpha H^{s+2}}{1+\frac{\alpha}{3} H^s}- 3 \zeta_0 H\right)
\label{dhdt}
\end{equation}
Using this in (\ref{dec3}), we get
 \begin{equation}
q(z)=-1+\frac{1}{2}\left(\frac{\alpha H^s}{1+\frac{\alpha}{3} H^s}- 3 \frac{\zeta_0}{H}\right)
\label{qofz}
\end{equation}
Solving for $\zeta_0$ at $z=0$ we get
\begin{equation}
\zeta_0=\frac{2 H_0}{3}\left[\left(-1-q_0\right)+\frac{1}{2}\left(\frac{\alpha {H_0}^s}{1+\frac{\alpha}{3} {H_0}^s}\right)\right]
\label{zeta}
\end{equation}
This equation constrains the bulk viscosity coefficient $\zeta_0$. If we are able to constrain the two parameters $\alpha$ and $s$, we can calculate $\zeta_0$ through this relation using the two observational values of $q_0$ and $H_0$.

\subsection{Constraining the Remaining two parameters} \label{othpar}
The remaining two parameters $\alpha$ and $s$ can also be constrained based upon other cosmological observations, such as cosmic deceleration-acceleration transition redshift and the age of the universe.

\subsubsection{Cosmic Deceleration-Acceleration Transition} \label{dectran}
Observations from type Ia Supernovae (SNe Ia) and cosmic microwave background (CMB) support the scenario of the current accelerated expansion of the universe due to the domination of dark energy budget, and a decelerated expansion of earlier times due to the domination of cold dark and baryonic matter. This means that the universe underwent a dynamical phase transition from deceleration to acceleration at some transition redshift $z_{tr}$. At this value of the redshift, the deceleration parameter is zero. Using the value of $z_{tr}$ due to observations in Eq.(\ref{qofz}), we can constrain one of the remaining two parameters, $\alpha$ or $s$.

\subsubsection{Age of the Universe} \label{age}
The age of the universe is a powerful tool for examining cosmological models and also for adjusting their parameters. The lower limit to the age of the universe is obtained by dating the oldest stellar populations. Of special interest in this regard are globular clusters, the oldest objects in our galaxy. Each cluster has a chemically homogeneous population of stars all born nearly simultaneously. 
There are three ways to reliably infer the age of the oldest stars in the galaxy\cite{kra}: radioactive dating, white dwarf cooling and the main sequence turnoff time scaling. A summary for the universe age estimates due to different models and measurements is given by Spergel et al. $(2003)$\cite{sprg}. The range $[11-16]$ Gyr is estimated due to globular clusters age. Radioactive dating estimates resulted in the range $[9.5-20]$ Gyr, while white dwarfs put a lower limit of $12.5\pm0.7$ Gyr. Krauss and Chaboyer $(2003)$\cite{kra} estimated the age of globular clusters using Monte Carlo simulation. They estimated a range of $[11-16]$ Gyr to the age of the universe. Kristiansen and Elgaroy\cite{kri} used a combination of cosmic microwave background (CMB), large scale structure (LSS), and SNe Ia data to get a lower limit of $12.58\pm0.26$ Gyr to the expansion age of the universe.

In this context, we'll use the recent data for the age of the universe to constrain the last parameter in our model. In fact, it is well known that the available data for the universe age is model dependent. However, to settle our model, it is important to ensure that it'll not suffer a cosmic age problem. Accordingly, at this step, we rely on the available data for the age of the universe to constrain one of the model parameters, and we'll then test the model against several model independent observations.

Theoretically, the age of the universe can be calculated through the relation
\begin{equation}
{\int_0}^{t_0} dt = - {\int_{H_0}}^\infty \frac{dH}{\dot{H}} 
\label{t0}
\end{equation}
Using (\ref{dhdt}) we get for our model
\begin{equation}
t_0= {\int_{H_0}}^\infty \frac{2\left(1+\frac{\alpha}{3} H^s\right)}{\alpha H^{s+2}- 3 \zeta_0 H \left(1+\frac{\alpha}{3} H^s\right)} dH 
\label{aget0}
\end{equation}
\bigskip

Proceeding this way, all of our parameters are now constrained. In our calculations the Hubble constant, $H_0$, is taken as $70$ $km$ $s^{-1}  Mpc^{-1}$ \cite{coll}, deceleration parameter as $q_0=-0.57$ \cite{varg}, deceleration acceleration transition redshift as $z_{tr}=0.76$ \cite{far}, and the age of the universe as $t_0=13.8$ Gyr \cite{plank}. The parameters of the model are constrained due to these observations to the values $s=3.92$, $\widetilde{\alpha} = 81.65$ $km$ $s^{-1}  Mpc^{-1}$, and $\widetilde{\zeta_0}=0.057$.

\section{Expectations for the Universe Evolution} \label{test}
In the following, we study the model expectations for the universe evolution. However, we first confront our model with some physical and cosmological observational tests.

\subsection{The $Om(z)$ Diagnostic} \label{digtest}
The behavior of the cosmological models is well defined through the behavior of their cosmological parameters such as the Hubble parameter, the deceleration parameter, and the EoS parameter. However, all acceptable cosmological models have a positive Hubble parameter $H(z)$, and a deceleration parameter $q(z)$ switches sign during evolution from positive to negative indicating a phase transition from deceleration to acceleration at a given value of $z$ known from observations. Accordingly, these two parameters cannot differentiate quintessence-like from phantom-like models. Even the effective EoS parameter for dynamical dark fluid models may not be enough to effectively differentiate models.

As $H$ is a function of $\dot a$ and $q$ is a function of $\ddot a$, one way to differentiate models is to use higher time derivatives for the scale parameter. There are two parameters that are functions of $\dddot{a}$ which are called statefinder parameters  \cite{alam, sah}, usually denoted as $\left\{r, s\right\}$. These two parameters do, in fact, differentiate models. However, these are functions of $\dddot{a}$, so they need somewhat heavy calculations. A more simple, while also effective, way to do the job is to apply the $Om(z)$ diagnostic test. It relies only on the first order derivative and so demands less effort.  

The $Om(z)$ diagnostic test was introduced by Sahni et al. $(2008)$ \cite{sah} to distinguish the behavior of DE in dynamical models away from the EoS.  His relation stems from the redshift dependence of the function $H^2$ and is given by
\begin{equation}
Om(z)=\frac{\left(H/{H_0}\right)^2-1}{\left(1 + z\right)^3 -1}
 \label{omofz}
\end{equation}

In unified models, the two dark sectors are treated as one entity. If the single fluid is split into two components, so that
\begin{eqnarray}
\rho(z)= \rho_{m}(z) + \rho_{de}(z)    \label{rhotot}
\end{eqnarray}
we then have
\begin{eqnarray}
 \rho(z)= \rho_{{m}_0} \left(1 + z\right)^3 + \rho_{de}(z)    \label{rhoto0}
\end{eqnarray}
Using (\ref{hsq}) and divide by the critical density, $\rho_c(z)$, we get
\begin{equation}
H^2={H_0}^2\left[\Omega_{{m}_0}\left(1 + z\right)^3 + \Omega_{{de}_0} g(z)\right]
 \label{gofz}
\end{equation}
where $g(z)$ is defined through the relation $\rho_{de}(z)=\rho_{{de}_0} g(z)$. It follows that
\begin{equation}
Om(z)=\frac{1}{\left(1 + z\right)^3 -1}\left[\Omega_{{m}_0}\left(1 + z\right)^3 + \Omega_{{de}_0} g(z)  -1\right] 
 \label{omofz1}
\end{equation}
For a spatially flat universe we then have
\begin{equation}
Om(z)=\Omega_{{m}_0} + \frac{\left(1-\Omega_{{m}_0}\right)\left(g(z)-1\right)}{\left(1 + z\right)^3 -1} 
 \label{omofz2}
\end{equation}

The $\Lambda$CDM model considers a cosmological constant $\Lambda$ for DE and hence has $g(z)=1$. Accordingly, it has $Om(z)=\Omega_{{m}_0}$, which means that $Om(z)$ is just a null test for the cosmological constant. On the other hand, for any other dynamical model, $Om(z)>\Omega_{{m}_0}$ represents quintessence, while $Om(z)<\Omega_{{m}_0}$ represents phantom.

	\begin{figure} [ht]
	\centering
		\includegraphics[width=8.5cm, height=8cm] {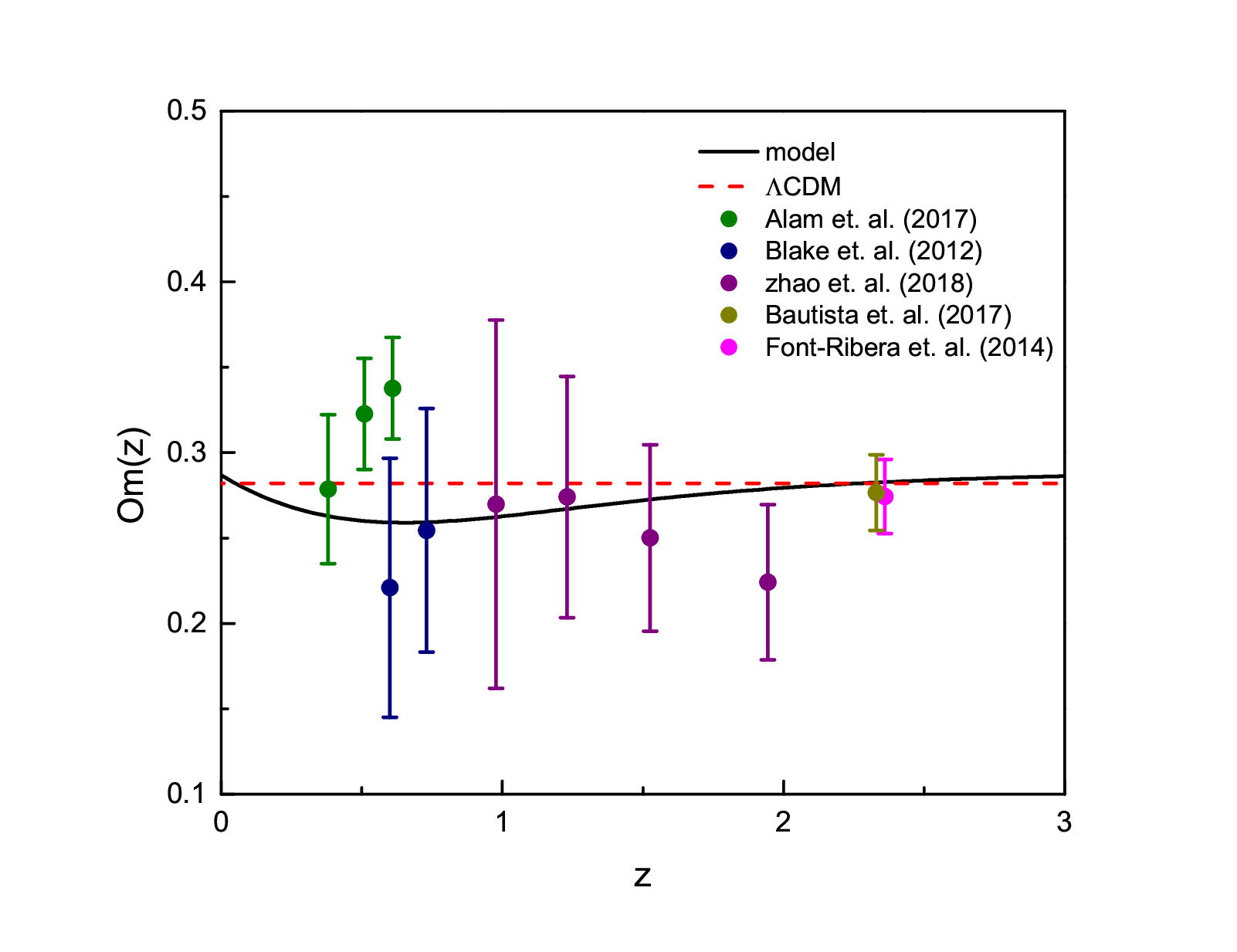}
  	\vspace{-.8cm}
	\caption{\scriptsize{Evolution of the $Om(z)$ parameter, solid curve. Shown also data from Alam et al. (2017) \cite{alam1}, Blake et al. (2012) \cite{blak}, Zhao et al. (2018) \cite{zhao}, Bautista et al. (2017) \cite{baut}, and Font-Ribara et al. (2014) \cite{font}. All error bars are $1\sigma$.}}
	\label {fig:omofz}
\end{figure}

In Fig. \ref{fig:omofz} we plot $Om(z)$ for our model together with the $\Lambda$CDM limit. We also show the BAO data points for this quantity. Data from Alam et al. (2017) \cite{alam1}, Blake et al. (2012) \cite{blak}, Zhao et al. (2018) \cite{zhao}, Bautista et al. (2017) \cite{baut}, and Font-Ribara et al. (2014) \cite{font} are shown. The Fig. shows that our model is consistent with the data points for the full redshift range. We refer to the standard $\chi^2$ statistics to measure the goodness of fit, where the $\chi^2$ value is calculated from
\begin{equation}
\chi^2=\sum_i{\frac{(Om_{th}(z_i)-Om_{obs}(z_i))^2}{\sigma^{2}_{i}}}
\label{chi}
\end{equation}
with $\sigma_i$ represents the error due to each $Om_{obs}$. However, due to the tension between Alam et al. and Blake et al. datasets, we consider the $\chi^2$ calculations including either of them at a time. Including Blake et al. dataset results in the value of a ${\chi^2}_{red}$ of $0.30$ with a $p-$value of $0.95$ for our model compared to ${\chi^2}_{red}=0.42$ with a $p-$value of $0.88$ for the $\Lambda$CDM model. On the other hand, including Alam et al. dataset instead results in the value of ${\chi^2}_{red}$ of $1.58$ with a $p-$value of $0.13$ for our model compared to ${\chi^2}_{red}=0.90$ with a $p-$value of $0.51$ for the $\Lambda$CDM model. This shows that our model prefers Blake et al. data other than Alam et al. one.

\subsection{The Hubble Parameter} \label{hubtest}
Hubble parameter, $H(z)$, is the parameter that measures the cosmological expansion rate. Observational $H(z)$ data are obtained from model independent direct measurements. Accordingly, they can be used to constrain conventional cosmological parameters such as $z_{tr}$, $\Omega_{m0}$, and $\Omega_{de0}$. Two major methods have been developed for measuring $H(z)$, galaxy differential age and radial BAO size methods \cite{zng}. Recently, Farooq et al. \cite{far} compiled an updated list of $38$ measurements of $H(z)$ for $0.07 \leq z \leq 2.36$.

In our model, we can solve for a relation between the Hubble parameter and redshift. Eq.(\ref{dhdt}) can be rewritten as
\begin{equation}
\frac{dH}{dz}=\frac{1}{2(1+z)}\left(\frac{\alpha H^{s+1}}{1+\frac{\alpha}{3} H^s}- 3 \zeta_0\right)
\label{dhdz}
\end{equation}
This can be solved numerically to get $H(z)$.
 \begin{figure} [ht]
	\centering
		\includegraphics[width=7cm, height=7cm] {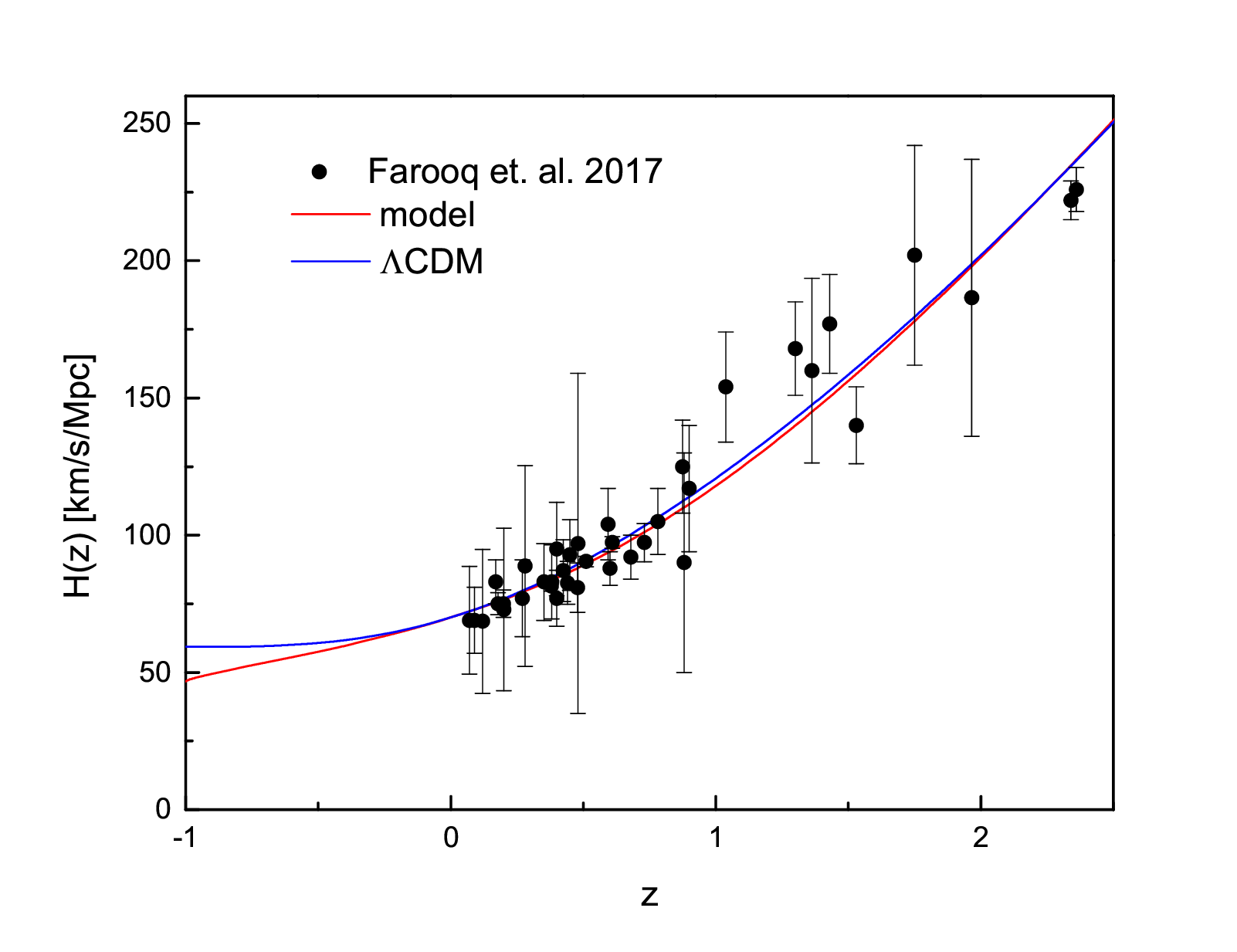}
  	\vspace{-.3cm}
	\caption{\scriptsize{Hubble parameter from our model compared to observations from Farooq et al. \cite{far}, where error bars are $1 \sigma$. Results from the $\Lambda$CDM model are also shown for comparison.}}
	\label {fig:hubble}
\end{figure}
In Fig. \ref{fig:hubble} we plot results from our model against the list of Farooq et al. \cite{far}. Results from the $\Lambda$CDM model are also shown for comparison. The Fig. shows that the model represents the data well and agrees at large $z$ with the results of the $\Lambda$CDM model. 

Referring to the $\chi^2$ statistics for the goodness of fit, we considered the $38$ observations of Farooq et al. against the numerical solution of (\ref{dhdz}) for our model. A ${\chi^2}_{red}$ value of $0.672$ is obtained for our model with the $p-$value of $0.94$. This reflects the capability of our model to efficiently fit the Hubble parameter data. It also shows that our model can cope with large values of the Hubble constant from local observations.

\subsection{Supernovae SNe Ia Observations} \label{super}
Pantheon Sample \cite{pan} is the latest compilation of Type Ia supernovae (SNe Ia). Confirmed SNe Ia from Pan-STARRS1 Medium Deep Survey are combined with previous available SNe Ia samples from other surveys to form a sample of $1048$ SNe Ia. Observations of the apparent magnitude of SNe Ia, $m(z)$, are related to the distance modulus $\mu(z)$ through the relation
\begin{equation}
\mu(z)=m(z)-M=5 log_{10}(d_L(z))+25
\label{mu}
\end{equation}
Here $M$ is the absolute magnitude, and $d_L$ is the luminosity distance measured in $Mpc$. Theoretically, $d_L$ is given by
\begin{equation}
d_L=c (1+z) \int^{z}_{0} \frac{d\zeta}{H(\zeta)}
\label{dl}
\end{equation}

\begin{figure} [ht]
	\centering
		\includegraphics[width=8.5cm, height=8cm] {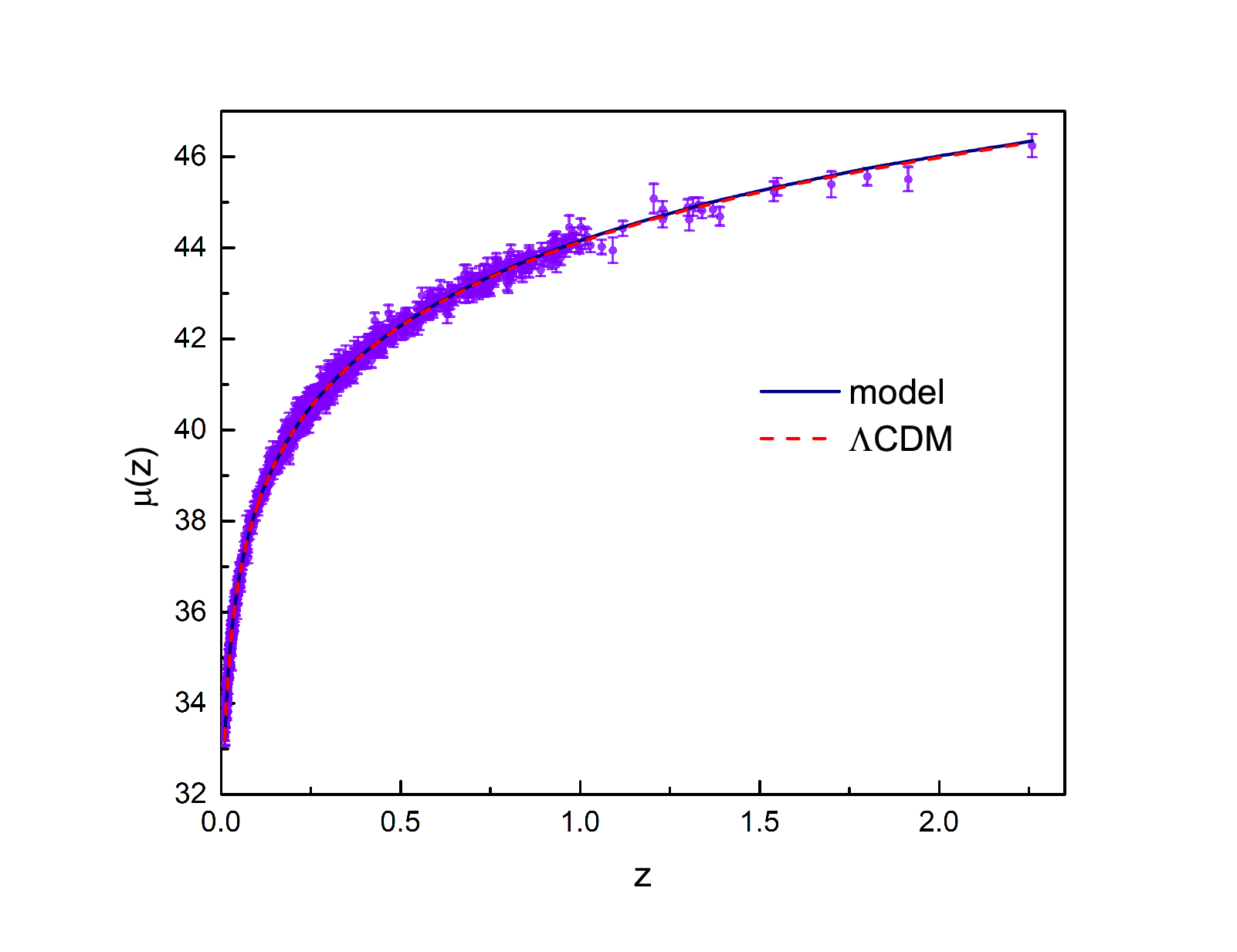}
		\vspace*{-.4cm}
	\caption{\scriptsize{Model versus observations for SNe Ia from Pantheon Sample. Results from $\Lambda$CDM are also shown for comparison.}}
	\label {mufig}
\end{figure}

For our model, the integration in the above equation can be calculated numerically. In Fig \ref{mufig}, results for the distance modulus are plotted against observations from Pantheon sample. Results for the $\Lambda$CDM scenario are also shown. 

The Fig. shows that our model matches SNe Ia observations excellently. In addition, It highly agrees with the results from the $\Lambda$CDM model. In the $\chi^2$ calculations, the summation of the $\chi^2$ relation
\begin{equation}
\chi^2=\sum_i{\frac{(\mu_{th}(z_i)-\mu_{obs}(z_i))^2}{\sigma^{2}_{i}}}
\label{chi1}
\end{equation}
is taken over the $1048$ data point, where $\sigma_i$ is the error due to each $\mu_{obs}(z_i)$. We obtained a ${\chi^2}_{red}$ value of $1.04$ with a $p-$value of $0.17$, which reflects the ability of our model to match SNe Ia data.

\subsection{Evolution of the Deceleration Parameter} \label{dtest}
Deceleration parameter (DP) is the parameter which was supposed to be a measure of deceleration of expansion of the universe due to gravity. However, cosmological observations of the high redshift supernovae in (1998) presented convincing evidence of the fact that the expansion of the universe is instead accelerated \cite{ries, perl}.

The redshift evolution of DP can be visualized using relation (\ref{ddot}) in eq. (\ref{dec1}). This gives
\begin{equation}
q(z)=\frac{1}{H^2} \frac{4 \pi G}{3} \left(\rho+3 \omega \rho\right)
 \label{qstan}
\end{equation}
Applying (\ref{hsq}) and considering units with $8 \pi G = 1$ we get
\begin{equation}
q(z)=\frac{1}{2} \left(1 + 3 \omega \right)
 \label{qst}
\end{equation}
We can see from the above relation that $q(z)$ changes from $+\frac{1}{2}$ for a matter dominated era to $-1$ for dark energy dominated era.

\begin{figure} [ht]
	\centering
		\includegraphics[width=7cm, height=7cm] {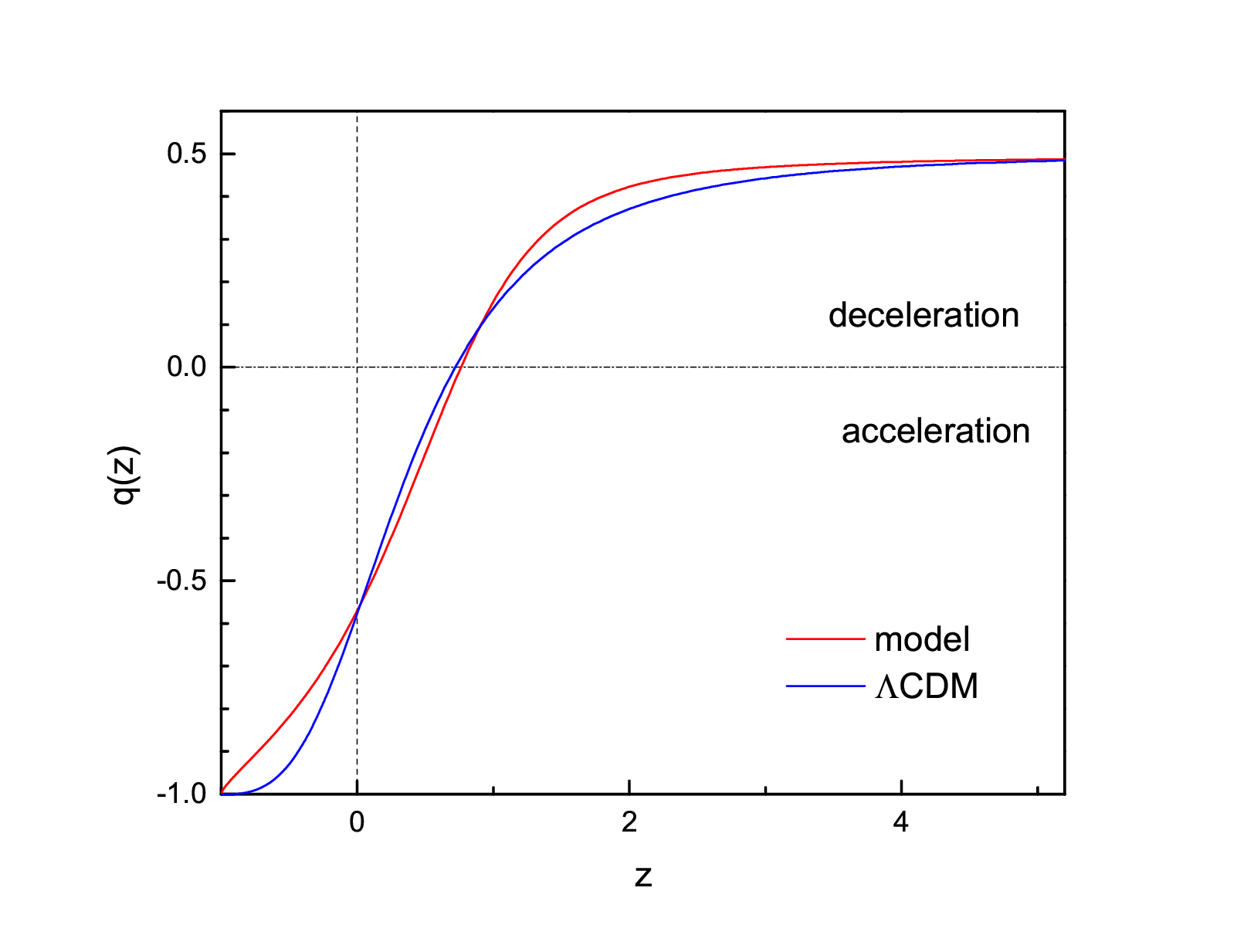}
  	\vspace{-.4cm}
	\caption{\scriptsize{Deceleration Parameter for our model compared to that of the $\Lambda$CDM model.}}
	\label {fig:dec}
\end{figure}

For our model, deceleration parameter can be calculated directly from relation (\ref{qofz}) using the numerical integration of relation (\ref{dhdz}). Fig. \ref{fig:dec} shows a plot for $q(z)$ for our model compared to the $\Lambda$CDM model.

It is clear from the Fig. that the deceleration parameter exhibits the behavior that is well known in the literature. We can also see the values of $q_0=-0.57$ and $z_{tr}=0.761$.

\subsection{Density of the Universe} \label{dentes} 
\begin{figure} [ht]
	\centering
		\includegraphics[width=7cm, height=7cm] {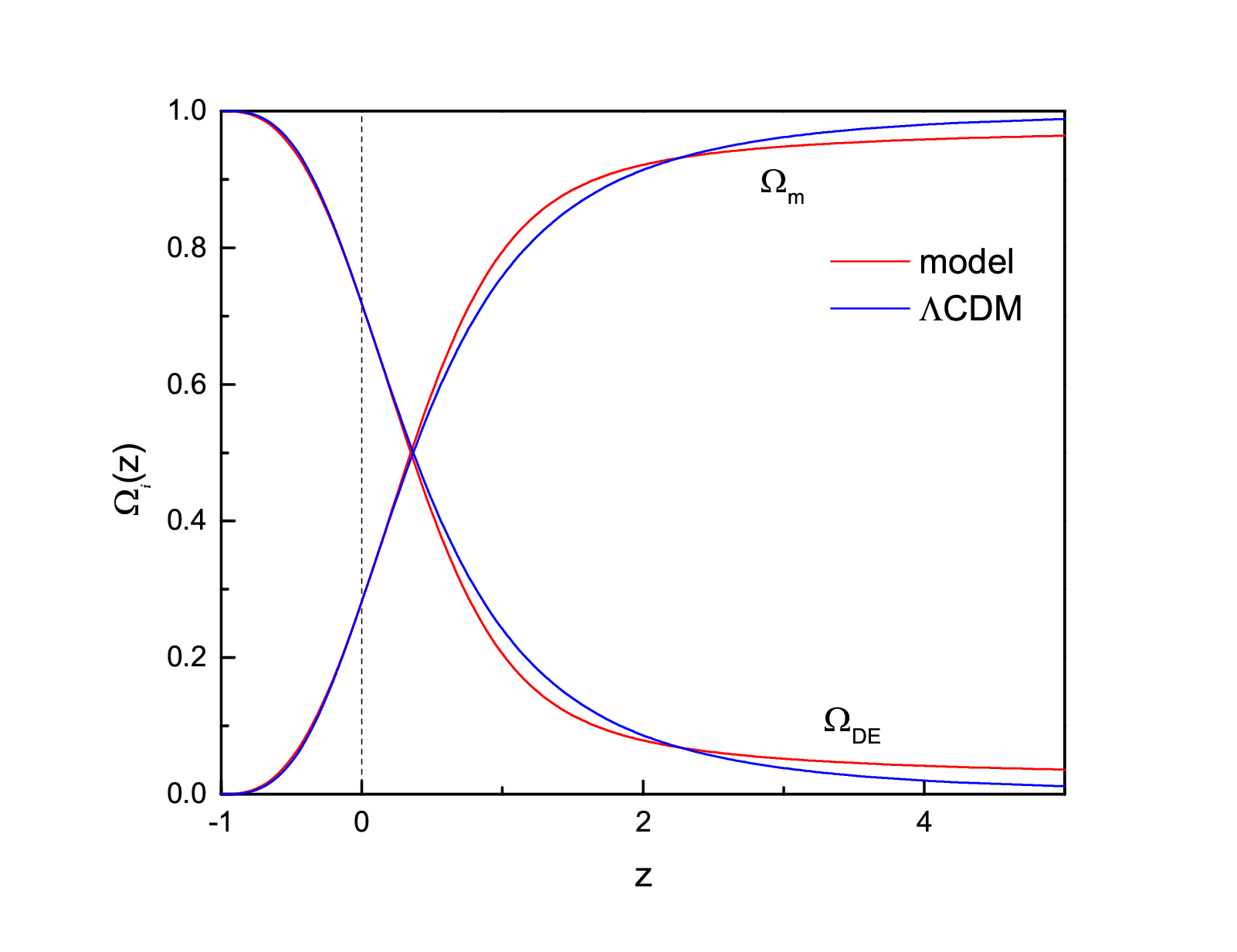}
  	\vspace*{-.4cm}
	\caption{\scriptsize{Matter and dark energy density parameters for our model compared to the $\Lambda$CDM model.}}
	\label {fig:densty}
\end{figure}

Referring to relations (\ref{rhotot})-(\ref{gofz}), the density parameters of the fluid candidates are given by
\begin{align}
\Omega_{m}(z)= \frac{{H_0}^2}{H^2} \Omega_{{m}_0}\left(1 + z\right)^3     \label{Omegam} \\
\Omega_{de}(z)= \frac{{H_0}^2}{H^2} \Omega_{{de}_0} g(z)     \label{Omegade}    
\end{align}

using the numerical integration of relation (\ref{dhdz}), the matter and dark energy density parameters can be graphically represented. The parameter $\Omega_{{m}_0}$ is taken as $0.282$ \cite{wmap}. Fig. \ref{fig:densty} shows $\Omega_i$ for each component due to our model compared to the $\Lambda$CDM model. 
The Fig. again represents the behavior that is well known in the literature. It is also clear that while our model has the same trend as the $\Lambda$CDM model at low redshift, it behaves differently at high redshift. Specifically, the dark energy density of our model decays more slowly than that of the $\Lambda$CDM model which reflects the smooth transition of our model. We can also see that the model expects a redshift of $z=0.35$ for both dark sectors to share half the energy density of the universe.

\subsection{Effective Equation of State Parameter} \label{effeq}
While the cosmological constant $\Lambda$ represents DE candidate for the $\Lambda$CDM model, dynamical models have effective time evolving DE candidate with a non-trivial parametrized equation of state.

Consider the splitting of the single fluid into two components. The effective EoS parameter of the cosmic fluid, ${\omega}_{eff}(z)$, is related to the parameter of the underlying dark energy EoS, $\omega_{de}(z)$, through  the relation
\begin{equation}
\omega_{de}(z)= \frac{{\omega }_{eff}(z)}{\Omega_{de}(z)}
\label{omgs1}
\end{equation}
The effective EoS parameter for the cosmic fluid, ${\omega }_{eff}(z)$, is calculated by considering relations(\ref{pres}) and (\ref{van1}). This gives
\begin{equation}
  {\omega}_{eff}(z)=-1 + \frac{\gamma \rho^{n-1}}{1+\delta \rho^m}- \frac{\zeta_0} {H}
\label{effp}
\end{equation}
Using (\ref{hsq}) and (\ref{const}) we get
\begin{equation}
 \omega_{eff}(z)=-1 + \frac{\alpha H^s}{3 + \alpha H^s} - \frac{\zeta_0}{H}
\label{wmod}
\end{equation}

On the other hand, the conservation equation (\ref{cons}) is written for dark energy candidate as
\begin{equation}
\dot{\rho_{de}}+3 \frac{\dot{a}}{a} (\rho_{de} + P_{de})=0
\label{decon}
\end{equation}
This can be written in the form
\begin{equation}
\dot{\rho_{de}}+3 \frac{\dot{a}}{a} (1 + \omega_{de}(a))\rho_{de}=0
\label{decon1}
\end{equation}
On using
\begin{equation}
\frac{da}{a}=-\frac{dz}{1+z}
\label{dadz}
\end{equation}
the integration of (\ref{decon1}) results in evolving dark energy density;
\begin{equation}
\rho_{de}(z)= \rho_{{de}_0} exp\left[3 {\int_0}^z \frac{1 + \omega_{de}(z)}{1+z} dz\right]
\label{rhode}
\end{equation}
Dividing by $\rho_c$ and substituting in (\ref{gofz}), we get 
\begin{equation}
H^2={H_0}^2 \left\{\Omega_{{m}_0}\left(1 + z\right)^3 + \Omega_{{de}_0} exp\left[3 {\int_0}^z \frac{1 + \omega_{de}(z)}{1+z} dz\right]\right\}
 \label{hsq1}
\end{equation}
which on comparison with Eq.(\ref{gofz}) gives
\begin{equation}
g(z)= exp\left[3 {\int_0}^z \frac{1 + \omega_{de}(z)}{1+z} dz\right]
\label{gofz1}
\end{equation}

\begin{figure} [ht]
	\centering
		\includegraphics[width=7cm, height=7cm] {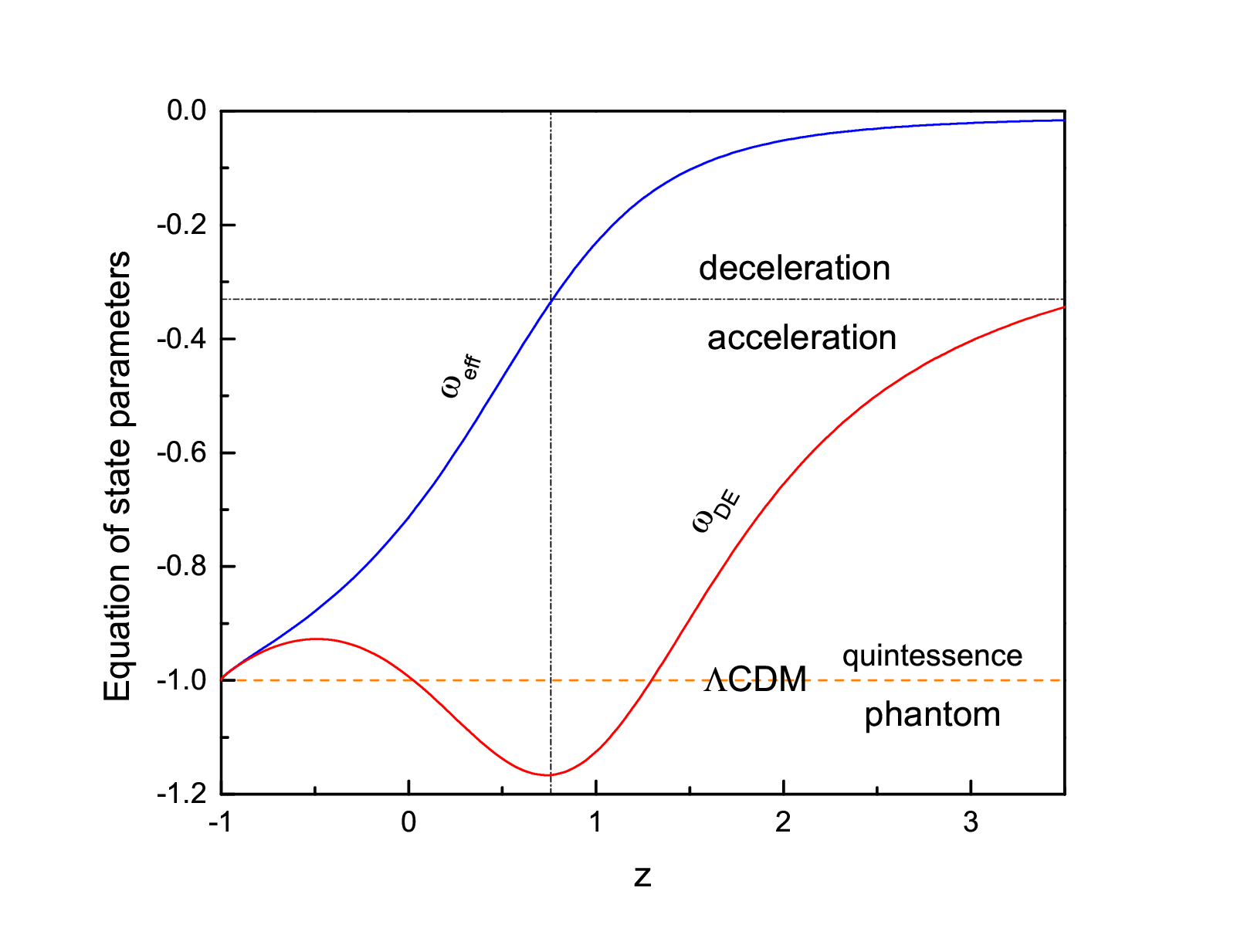}
  	\vspace{-.4cm}
	\caption{\scriptsize{Evolution of the effective EoS parameters for the cosmic fluid and the underlying DE budget.}}
	\label {fig:weff}
\end{figure}

Relations (\ref{omgs1}), (\ref{wmod}), and (\ref{hsq1}) can be used to study the behavior of the effective EoS parameter for the cosmic fluid, in addition to its underlying dark energy candidate.  Fig. \ref{fig:weff} shows a plot of such parameters.

The Fig. shows that the effective EoS of the cosmic fluid follows the dust EoS at large $z$. It drops to $\omega_{eff}(z=0)=-0.713$ at the present time, while tends to the value of $-1$ as $z \rightarrow -1$, whence represents a universe dominated by DE. The universe effectively crosses the value of $\omega_{eff}=-1/3$ at the transition redshift $z=0.76$ in agreement with expectations for the transition time. On the other hand, the DE parameter exhibits the current value of $\omega_{de}(z=0)=-0.99$. Moreover, it tends to $-1$ in the future, $\omega_{de}\rightarrow -1$ when $z\rightarrow -1$, in agreement with the expectations that the universe will be fully dominated by DE.

\subsection{Estimating the Viscosity of the Cosmic Fluid}
Due to Velten and Schwarz \cite{velt}, viscus dark matter allowed to have a bulk viscosity $\leq 10^7$ $Pa$ $s$. On the other hand, Wang and Meng \cite{wang} considered a $\Lambda$CDM model with time dependent bulk viscosity coefficient, $\zeta=\zeta(t)$. Their analysis resulted in a current bulk viscosity given in the average by $\zeta_0 \approx 10^5$ $Pa$ $s$, in agreement with Velten and Schwarz. Another work is done with Sasidharan and Mathew \cite{sasi} in which they performed a phase space analysis of a universe dominated by a bulk viscous matter. They considered a bulk viscosity coefficient which is a function of Hubble parameter and its first derivative. Their $\chi^2$ analysis to Supernovae data for the constant bulk viscosity coefficient model resulted in a current bulk viscosity of $\zeta_0 \approx 7.68 \times 10^7$ $Pa$ $s$.

Our model considers a viscous unified dark fluid. In Sec. \ref{cosmpar} we constrained the parameter $\widetilde{\zeta_0}$ based upon observations. Its estimated value is $0.057$. Referring to relation (\ref{tzeta}), the viscosity coefficient $\zeta_0$ has then the value of $4.68$ $km$ $s^{-1}  Mpc^{-1}$. Multiplying this by $c^2/ 8 \pi G $, the cosmic fluid is found to possess a bulk viscosity of $8\times10^6$ $Pa$ $s$.

\section{Conclusions} \label{conc}
A dissipative unified dark fluid model is considered. The EoS of the cosmic fluid can asymptote between two power laws so that it has the ability of smooth transition between dust and DE equations of state. The dissipative effects are added as a natural candidate of the real fluid. These dissipative effects are described by bulk viscosity with constant coefficient. The shear viscosity is excluded due to the isotropy of the universe. The model is analyzed using the theory of dynamical systems. The phase portrait of the model showed that it has three solution classes. Two of these solutions contradict observational constraints of our universe. The third solution with positive $H$ and negative $\dot H$ matches the dynamics of our universe, and, due to the work of Awad \cite{aawad}, it is free from future finite time singularities of types $I$, $II$, and $III$. We then adopted such a solution.

The parameters of the model are then constrained. The behavior of the model at the early time showed that two out of the five parameters are not independent, we chose them to be $\beta$ and $r$. Accordingly, only three parameters remain free, $\alpha$, $s$, and $\zeta_0$. We constrain these parameters using many observational constraints: today's value of deceleration parameter, the redshift value of deceleration-acceleration transition, and the age of the universe. 

Later on, the model is investigated against several physical and cosmological observational tests. We began by considering the model independent $Om(z)$ diagnostic test. Results for $Om(z)$ are plotted in Fig. \ref{fig:omofz} against the BAO data for this quantity from different authors. The Fig. showed that the model is consistent with the data points for the full redshift range. Due to the apparent tension between data points of Alam et al. and Blake et al., we considered the calculation of the $\chi^2$ value including either of them at a time. Considering Blake et al. dataset results in a ${\chi^2}_{red}$ value of $0.30$ with a $p-$value of $0.95$ for our model compared to a ${\chi^2}_{red}$ value of $0.42$ with a $p-$value of $0.88$ for the $\Lambda$CDM model. Instead, considering Alam et al. data points  results in a ${\chi^2}_{red}$ value of $1.58$ with a $p-$value of $0.13$ for our model compared to a ${\chi^2}_{red}$ value of $0.90$ with a $p-$value of $0.51$ for the $\Lambda$CDM model. This shows that our model prefers Blake el al. data other than Alam et al. one. It also shows that our model is compatible with local measurements of Hubble constant.

The Hubble parameter equation is then solved numerically and results are plotted against the newly compiled data of Farooq et al. \cite{far} in Fig. \ref{fig:hubble}. The Fig. showed that the model represents the observational data well and agrees with $\Lambda$CDM model at large $z$. The $\chi^2$ test which considered the results of our model against the $38$ observations of Farooq et al. resulted in the value of ${\chi^2}_{red}=0.672$ for our model with the $p-$value of $0.94$. This reflects the capability of our model to fit Hubble parameter data.

The model is then investigated using SNe Ia observations. Results for the distance modulus due to the model are plotted in Fig \ref{mufig} against the Pantheon Sample which is the latest SNe Ia sample. In the same graph results for the $\Lambda$CDM scenario are shown for comparison. The Fig. showed that our model highly matches the observational data and coincides with the $\Lambda$CDM curve. We also calculated the $\chi^2$ value which was found to be ${\chi^2}_{red}=1.04$ with the $p-$value of $0.17$. This shows that our model is efficiently able to fit the SNe Ia observations. 

Evolution of the deceleration parameter is also studied. Its graphical representation reflected the behavior that is well known in the literature. The density parameters for the fluid components are calculated and plotted as functions of the redshift. The graphical representation again reflected the well known behavior in the literature. Moreover, it reflected the ability of our model to asymptote smoothly between the two extreme ends, pure matter and pure DE, as the DE density of our model decays more slowly than that of the $\Lambda$CDM model at large $z$. Matter-DE equality is found to occur at the redshift value of $z=0.35$.

Evolution of the effective EoS parameter of the cosmic fluid, $\omega_{eff}(z)$, is also studied together with the underlying DE EoS parameter, $\omega_{de}(z)$. The graphical representation of these parameters showed many important features. It manifested the fact that the effective EoS of the cosmic fluid follows the dust EoS at large $z$ and the DE EoS to the future. It also showed that the current value of the effective EoS parameter is $\omega_{eff}(z=0)=-0.713$. Another important note is that the universe effectively crosses the value of $\omega_{eff}=-1/3$ at the transition redshift $z=0.76$ in agreement with the expectations for the transition time. On the other hand, the representation of the DE EoS showed that its parameter has a current value of $\omega_{de}(z=0)=-0.99$. It also tends to the value of $-1$, $\omega_{de}\rightarrow -1$, in the future in agreement with the expectations that the universe will be fully dominated by DE. Finally, the value of the viscosity coefficient of the cosmic fluid is estimated based on observational constraints to be $8\times10^6$ $Pa$ $s$.

We conclude that our model could pass very important cosmological and observational tests. Graphical representations of different physical quantities such as $Om(z)$ and the effective EoS parameter showed that this model is different from the $\Lambda$CDM model, but it has the ability to fit the observations in different known regions. Our model is then able to describe the behavior of the universe evolution and efficiently agrees with observations.

\section{Aknowledgments} \label{akn}
The author would like to thank Dr. Adel Awad and Dr. Waleed El Hanafy for valuable discussions and guidance.

\end{document}